\documentclass[11pt]{article}
\usepackage{moriond,epsfig}

\def\be{\begin{equation}}
\def\ee{\end{equation}}
\def\bea{\begin{eqnarray}}
\def\eea{\end{eqnarray}}

\newcommand{\comm}[1]{\ensuremath{\mbox{\textup{#1}}}}
\newcommand{\subs}[1]{\ensuremath{\scriptstyle \mathit{#1}}}

\renewcommand{\arraystretch}{1.3}
\newcommand{\FCal}[1][]{\ifthenelse{\equal{#1}{1}}{\textsc{FCal1}}%
                         {\ifthenelse{\equal{#1}{2}}{\textsc{FCal2}}%
                           {\ifthenelse{\equal{#1}{3}}{\textsc{FCal3}}%
                                                  {\textsc{FCal}}%
                           }%
                         }%
                       }%

\newcommand{\alphas}{\ensuremath{\alpha_{s}}}

\newcommand{\dzero}{\ensuremath{\mbox{D\O}}}

\newcommand{\mycs}{\ensuremath{d^{\,2}\sigma/(d\et d\eta)}}
\newcommand{\mycsav}{\ensuremath{\langle \mycs \rangle}}

\newcommand{\z}{\ensuremath{z}}

\newcommand{\et}{\ensuremath{E_{T}}}

\newcommand{\peta}{\ensuremath{\eta}}				
\newcommand{\aeta}{\ensuremath{|\eta|}}				
\newcommand{\ipb}{pb$^{-1}$}

\newcommand{\met}{\mbox{${\hbox{$E$\kern-0.63em\lower-.18ex\hbox{/}}}_{T}\,$}}
\newcommand{\metvec}{\mbox{${\hbox{$\vec{E}$\kern-0.63em\lower-.18ex\hbox{/}}}_{T}\,$}}
\newcommand{\metx}{\mbox{${\hbox{$E$\kern-0.63em\lower-.18ex\hbox{/}}}_{x}\,$}}
\newcommand{\mety}{\mbox{${\hbox{$E$\kern-0.63em\lower-.18ex\hbox{/}}}_{y}\,$}}

\newcommand{\etaphi}{\ensuremath{\eta-\varphi}}

\newcommand{\Rsep}{\ensuremath{\mathcal{R}_{\subs{sep}}}}
\newcommand{\etal}{{\it et al.}}
\newcommand{\ppbar}{\ensuremath{p\overline{p}}}

\newcommand{\MRSTGU}{MRSTg$\uparrow$}
\newcommand{\MRSTGD}{MRSTg$\downarrow$}
\newcommand{\chisqdef}{\ensuremath{\displaystyle\mathrm{ \chi^{2}=\sum_{i\,,\,j}(D_{i}-T_{i})\times
\left[\comm{Cov}_{i\,,\,j}^{\comm{\footnotesize{full}}}\right]^{-1}\!\!\!\!\times(D_{j}-T_{j}) }}}
\newcommand{\covmtxdef}{\ensuremath{\displaystyle\mathrm{Cov_{i\,,\,j}^{full}=\sum_{\beta}^{errors}
\rho_{i\,,\,j}^{\beta}\times\sigma_{i}^{\beta}\times\sigma_{j}^{\beta}}}}

\begin{document}
\vspace*{4cm}
\title{CENTRAL AND FORWARD INCLUSIVE JETS AT THE TEVATRON}

\author{ L. BABUKHADIA }

\address{Department of Physics, State University of New York\\
         Stony Brook, NY 11794, USA\\
	 E-mail: blevan@fnal.gov}

\maketitle\abstracts{
  We report on a new measurement of the rapidity dependence of the 
  inclusive jet production cross section in \ppbar\ collisions at
  $\sqrt{s}=1.8$ TeV using 92 \ipb\ of data collected by the \dzero\ 
  detector at the Tevatron collider. The differential cross 
  sections, \mycsav, are presented as a function of jet transverse energy
  (\et) in five pseudorapidity (\peta) intervals, up to $\aeta=3$, 
  significantly extending previous CDF and \dzero\ measurements 
  beyond $\aeta=0.7$. The extended range of the measurement should 
  provide greater discrimination among different parton distribution 
  functions. We also discuss previous measurements of the inclusive jet 
  cross sections made by the two collider experiments 
  at central pseudorapidities up to $\aeta=0.7$. Finally, we present recent
  measurements from the CDF and \dzero\ experiments of the ratio of central 
  inclusive cross sections from two center-of-mass energies, $0.63$ TeV 
  and $1.8$ TeV, as a function of jet $x_{T}$. Experimental results are 
  compared to next-to-leading order QCD predictions.}

\section{Jet Cross Sections --- Tests of QCD}

\begin{figure}[!t] \centering
  \begin{picture}(160,65)  

  \put(-1,-2.3){\begin{picture}(70,65)
               \epsfxsize=6.7cm
               \epsfysize=7.17cm
               \epsfbox{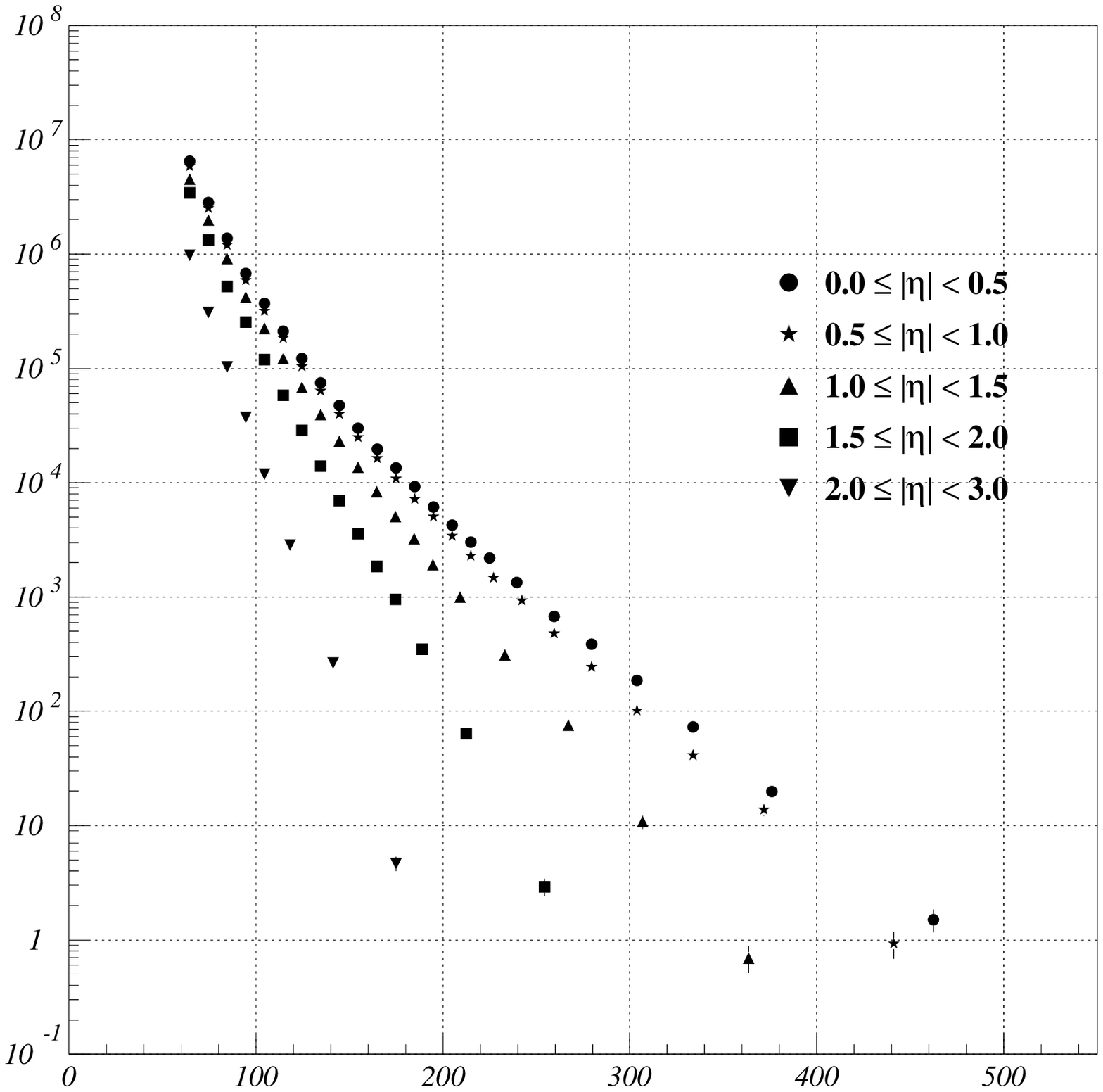}
            \end{picture}}
  \put(11.5,62.8)
    {\makebox(0,0)[rb]{\scalebox{0.6}{ \bf (a) }}}
  \put(-1.9,18){
    \scalebox{0.6}{ \rotatebox{90}{\boldmath $\mycsav\ \mathrm{(fb/GeV)}$}} }
  \put(61,0)
    {\makebox(0,0)[rb]{
      \scalebox{0.6}{ \boldmath  $\et\ \mathrm{(GeV)}$}}}
  \put(56,54)
    {\makebox(0,0)[rt]{
       \scalebox{0.5}{{\bf \dzero\ Run~1B \boldmath $\int \! \mathcal{L} dt = 92$ \ipb}}}}

  \renewcommand{\arraystretch}{1.0}
  \put(10,6.5)
    {\makebox(0,0)[lb]{\scalebox{0.4}{
       {\begin{tabular}{c}
           \bf nominal cross sections \& \\
           \bf statistical errors only \\
        \end{tabular}}}}}
  \renewcommand{\arraystretch}{1.3}

  \put(48,-16.0){\begin{picture}(70,75)
               \epsfxsize=8.1cm
               \epsfysize=9.65cm
               \epsfbox{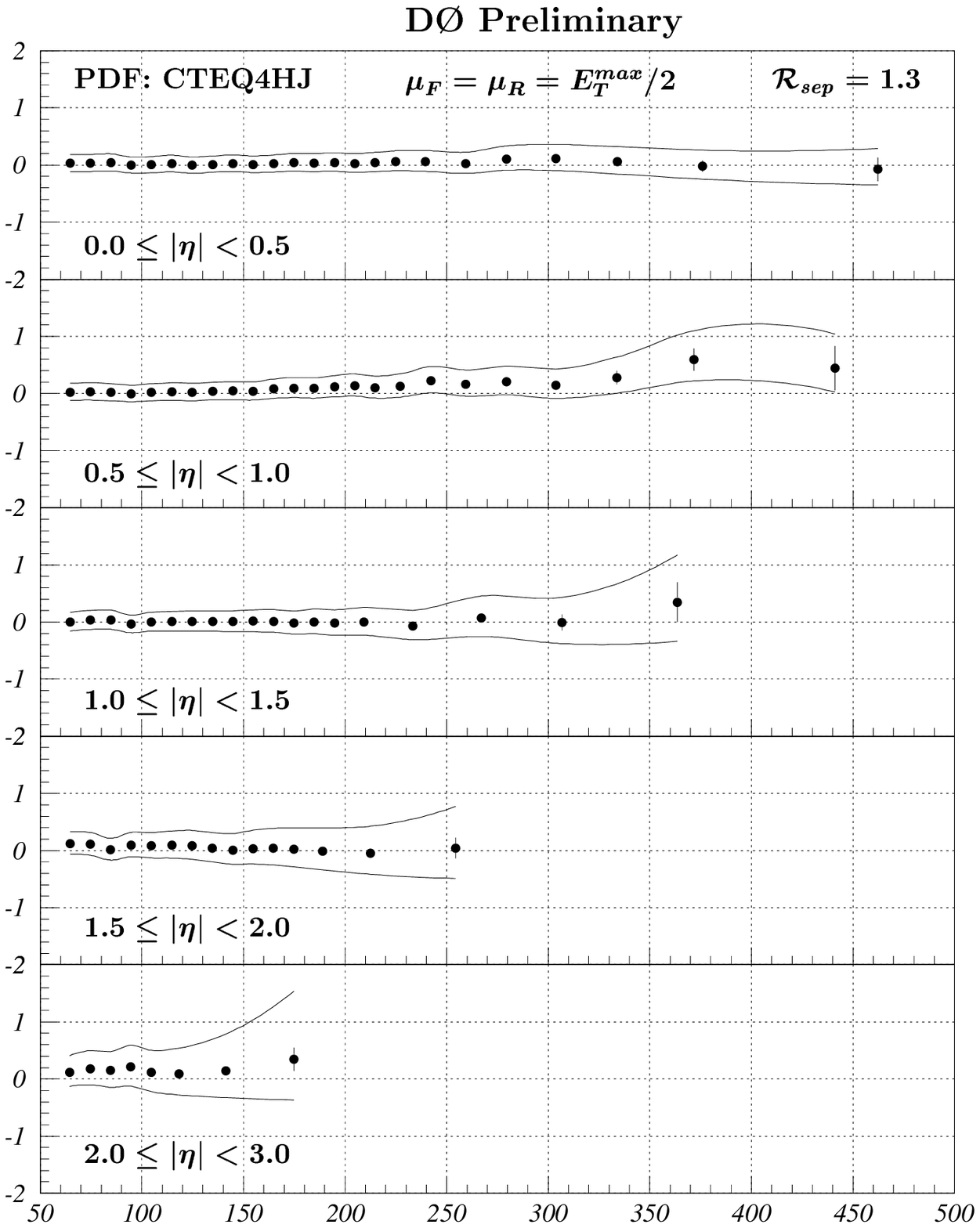}
             \end{picture}}
  \put(59.6,17.6){
    \scalebox{0.6}{ \rotatebox{90}{\bf ( Data - Theory ) / Theory} } }
  \put(117.5,0)
    {\makebox(0,0)[rb]{
      \scalebox{0.6}{ \boldmath  $\et\ \mathrm{(GeV)}$}}}
  \put(72,62.8)
    {\makebox(0,0)[rb]{\scalebox{0.6}{ \bf (b) }}}
  \put(99,53.1)
    {\makebox(0,0)[rt]{
       \scalebox{0.45}{{\bf JETRAD }}}}

  \renewcommand{\arraystretch}{2.40}
  \put(117.6,33.7){
        \scalebox{0.49}{
	\begin{tabular}{lcc} 
          \multicolumn{3}{c}{\boldmath \chisqdef} \\ 
          \multicolumn{3}{c}{\boldmath \covmtxdef} \\ \hline \hline
          PDF     & $0.0\leq\aeta<0.5$ & $0.1\leq\aeta<0.7$ \\ \hline
	  CTEQ3M  & 26.1 (35\%)        & 33.2 (10\%)        \\ 
	  CTEQ4M  & 20.5 (67\%)        & 27.2 (30\%)        \\ 
	  CTEQ4HJ & 16.6 (86\%)        & 22.3 (56\%)        \\ 
	  MRSA'   & 20.8 (65\%)        & 28.8 (23\%)        \\ 
	  MRST    & 25.2 (40\%)        & 29.5 (20\%)        \\ 
	  \MRSTGU & 21.5 (61\%)        & 30.1 (18\%)        \\ 
	  \MRSTGD & 47.3 (0.003\%)     & 47.5 (0.003\%)     \\ \hline\hline
	\end{tabular}
	}
	}
  \renewcommand{\arraystretch}{1.3}
  \put(123,62.8)
    {\makebox(0,0)[rb]{\scalebox{0.6}{ \bf (c) }}}

  %

  
  \end{picture}
  \caption{\dzero\ measurement of rapidity dependence of single inclusive jet 
           production cross section presented as a function of jet \et\ in five 
           jet \peta\ intervals (a), comparisons with $\alphas^{3}$ QCD 
           predictions calculated by JETRAD with CTEQ4HJ PDF (b). Table 
	   (c) shows $\chi^{2}$ values and corresponding probabilities for 
	   $24$ degrees of freedom for the previous \dzero\ measurement of the
	   inclusive jet cross section in two central intervals of pseudorapidity.}
  \label{fig:rapdep}
  \vspace{0.2cm}
\end{figure}

\noindent
In the last decade of the $20$th century, high energy physics saw 
impressive progress made in both theoretical and experimental 
understanding of collimated streams of particles or ``jets'' 
resulting from inelastic hadron collisions.
The Fermilab Tevatron \ppbar\ Collider, operated at 
center-of-mass energies of $0.63$ TeV and $1.8$ TeV, has been
a prominent arena for studying hadronic jets.
Theoretically, jet production in \ppbar\ collisions is 
understood within the framework of quantum chromodynamics
(QCD) as a hard scattering of constituents of protons, 
the quarks and gluons (or partons) that
manifest themselves as jets in the final state.
Studying various jet cross sections in 
CDF and \dzero, therefore provides stringent 
tests of QCD.

Perturbative QCD calculations of jet cross sections~\cite{theory},
using new and accurately determined parton distribution functions 
(PDFs)~\cite{pdfs}, add particular interest to the corresponding 
measurements at the Tevatron.
These measurements test the short range behavior of QCD, the 
structure of the proton in terms of PDFs, and any possible
substructure of quarks and gluons.
The measurements we report are based on integrated luminosities 
of $87$ and $92$ \ipb\ collected by the CDF and \dzero\ experiments, 
respectively, during the $1994$--$95$ Tevatron run.
In both experiments, jets are reconstructed using an iterative 
cone algorithm with a fixed cone radius of $\mathcal{R}=0.7$ in 
\etaphi\ space, where the pseudorapidity \peta\ is related to the 
polar angle (from the beam line) \mbox{$\theta$ via} 
\mbox{$\peta=\ln[\cot\theta/2]$} and $\varphi$ is the azimuth.
Offline data selections eliminate contamination from background 
caused by electrons, photons, noise, or cosmic rays.
This is achieved by applying an acceptance cutoff on the 
\z--coordinate of the interaction vertex, flagging events with large 
missing transverse energy, and applying jet quality criteria.
Details of data selection and corrections due to noise and/or 
contamination are described elsewhere~\cite{selection,mythesis}.
A correction for jet energy scale accounts for instrumental effects 
associated with calorimeter response, showering and noise, as well as 
for contributions from spectator partons, and corrects on average jets 
from their reconstructed to their ``true'' \et.
The effect of calorimeter resolution on jet cross section is removed 
through an unfolding procedure.
In \dzero, the energy scale and resolution corrections are determined 
mostly from data, and applied in two separate steps, while the CDF 
corrections are implemented in a single step by means of a Monte Carlo 
tuned to their data.

\section{Inclusive Jet Cross Sections at $\sqrt{s}=1.8$ TeV}
\noindent
\dzero\ has recently completed a measurement of the rapidity dependence 
of the inclusive jet production cross section~\cite{mythesis}.
The differential cross section, \mycsav, is determined as a function of 
jet transverse energy in five intervals of \aeta, up to $\aeta=3$, 
thereby significantly extending previously available measurements from 
CDF and \dzero\ beyond $\aeta=0.7$.
The cross section is calculated from the number of jets in each 
$\eta$--\et\ bin, scaled by the integrated luminosity, selection 
efficiencies, and the unfolding correction.
The measurement in each of the five \aeta\ regions is presented in 
Fig.~\ref{fig:rapdep}a.
The measurement spans about seven orders of magnitude in \et, and 
extends to the highest energies ever reached.

The results are compared to the $\alphas^{3}$ predictions from JETRAD 
(Giele, \etal~\cite{theory}), with equal renormalization and factorization 
scales set to $\et^{max}/2$, and using the parton clustering parameter 
$\Rsep=1.3$.
Comparisons have been made using all recent PDFs of the CTEQ and MRST 
families.
Figure~\ref{fig:rapdep}b shows the comparisons on a linear scale with 
the CTEQ4HJ PDF, which appears to best describe the data in all \peta\ 
intervals.
The error bars are statistical, while the error bands indicate $1$
standard deviation systematic uncertainties.
Theoretical uncertainties are on the order of the systematic errors.
Work is currently underway to obtain a more quantitative comparison 
with predictions (such as a $\chi^{2}$ test), taking into consideration 
correlations in \et\ and in \peta.
The extended range of the measurement promises to provide greater 
discrimination among different PDFs. 

\begin{figure}[!t] \centering
  \begin{picture}(160,65)   

  \put(0,0){\begin{picture}(70,75)
               \epsfxsize=5.3cm
               \epsfysize=6.1cm
               \epsfbox{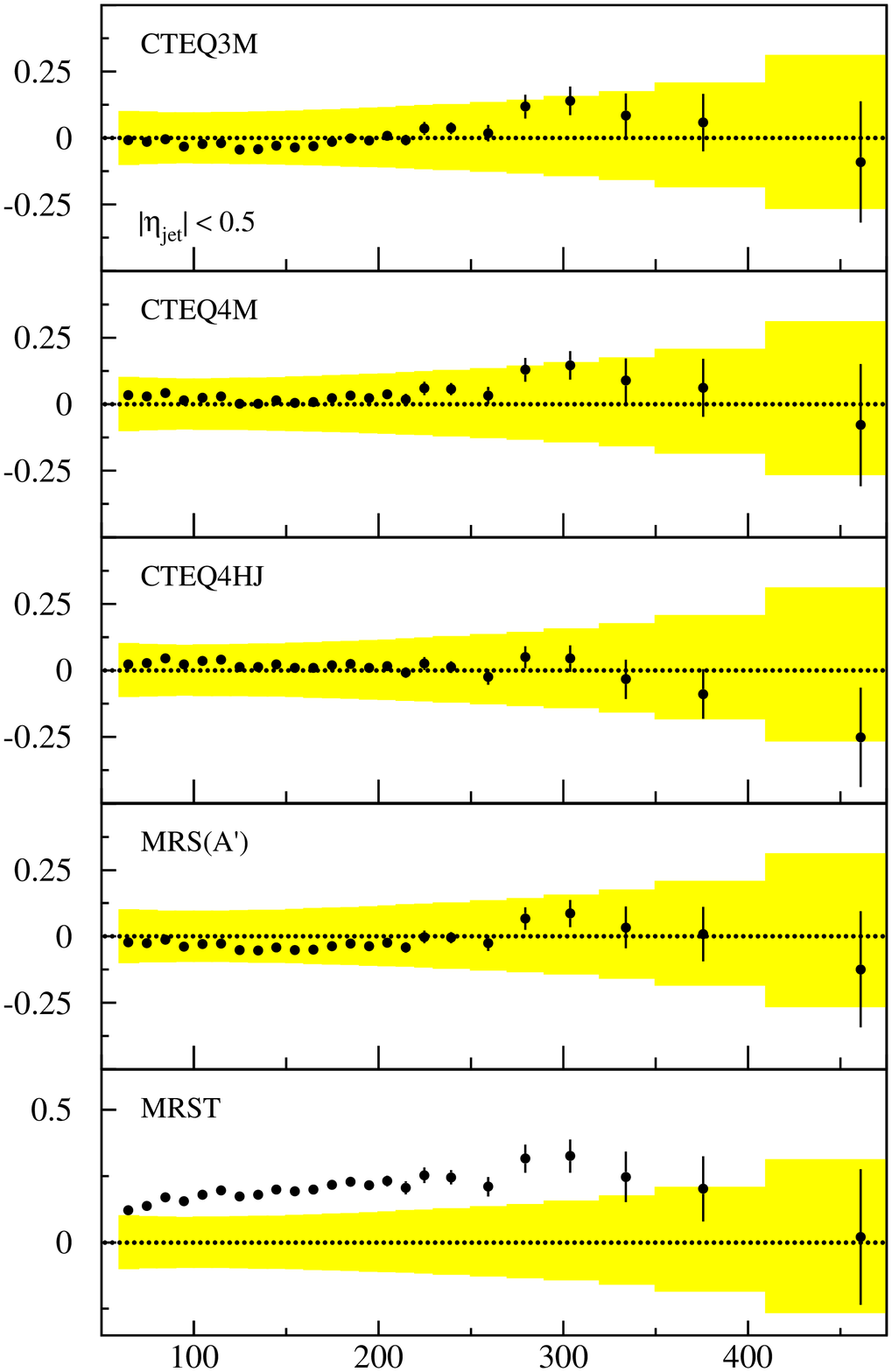}
             \end{picture}}
  \put(10,61)
    {\makebox(0,0)[rb]{\scalebox{0.6}{ \bf (a) }}}
  \put(-1.8,18){
    \scalebox{0.6}{ \rotatebox{90}{\bf ( Data - Theory ) / Theory} } }
  \put(52,0)
    {\makebox(0,0)[rb]{
      \scalebox{0.6}{ \boldmath  $\et\ \mathrm{(GeV)}$}}}
  \put(10,65)
    {\makebox(0,0)[lt]{
      \scalebox{0.6}{ \bf \dzero\ Inclusive Jet Cross Section }}}
  \put(18,59.8)
    {\makebox(0,0)[lt]{
      \scalebox{0.4}{ \bf JETRAD \boldmath $\;\;\mu_{R}=\mu_{F}=\et^{max}/2$ }}}
  \put(18,58)
    {\makebox(0,0)[lt]{
      \scalebox{0.4}{ \boldmath $\Rsep=1.3$ }}}

  \put(54,0.5){\begin{picture}(70,75)
               \epsfxsize=5.3cm
               \epsfysize=6.45cm
               \epsfbox{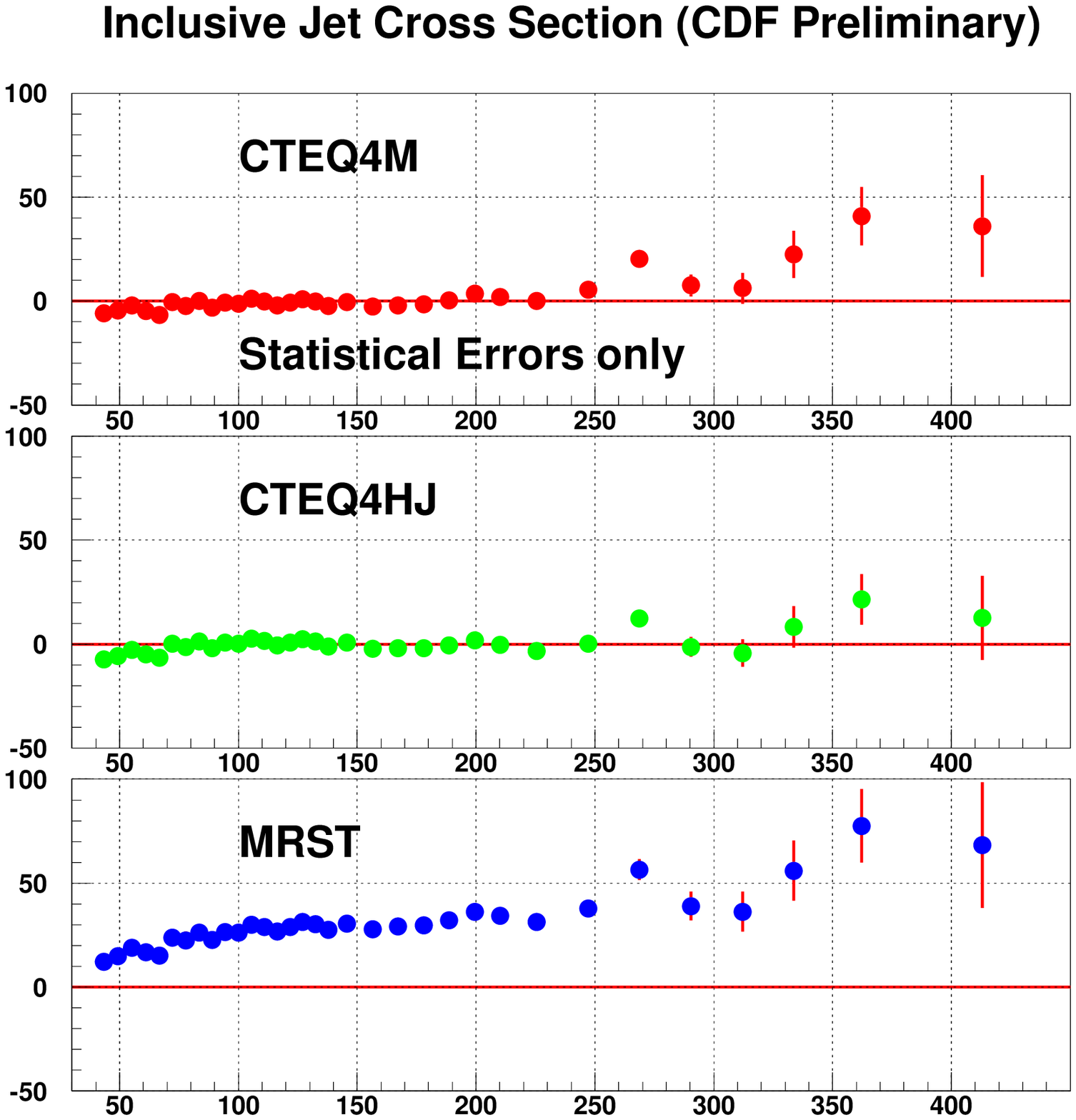}
            \end{picture}}
  \put(61,61)
    {\makebox(0,0)[rb]{\scalebox{0.6}{ \bf (b) }}}
  \put(51,18){
    \scalebox{0.6}{ \rotatebox{90}{\bf ( Data - Theory ) / Theory} } }
  \put(107.5,0)
    {\makebox(0,0)[rb]{
      \scalebox{0.6}{ \boldmath  $\et\ \mathrm{(GeV)}$}}}
  \put(82,59.5)
    {\makebox(0,0)[lt]{
      \scalebox{0.5}{ \bf EKS \boldmath $\;\;\mu_{R}=\mu_{F}=\et/2$ }}}
  \put(82,56.8)
    {\makebox(0,0)[lt]{
      \scalebox{0.5}{ \boldmath $\Rsep=1.3$ }}}
  \put(65.5,53.5)
    {\makebox(0,0)[lt]{
      \scalebox{0.5}{ \boldmath $0.1\leq\aeta<0.7$ }}}
  \put(106.3,46.6)
    {\makebox(0,0)[rt]{
       \scalebox{0.5}{{\boldmath $\int \! \mathcal{L} dt = 87$ \ipb}}}}

  \put(101,-1){\begin{picture}(70,75)
               \epsfxsize=6.6cm
               \epsfysize=6.2cm
               \epsfbox{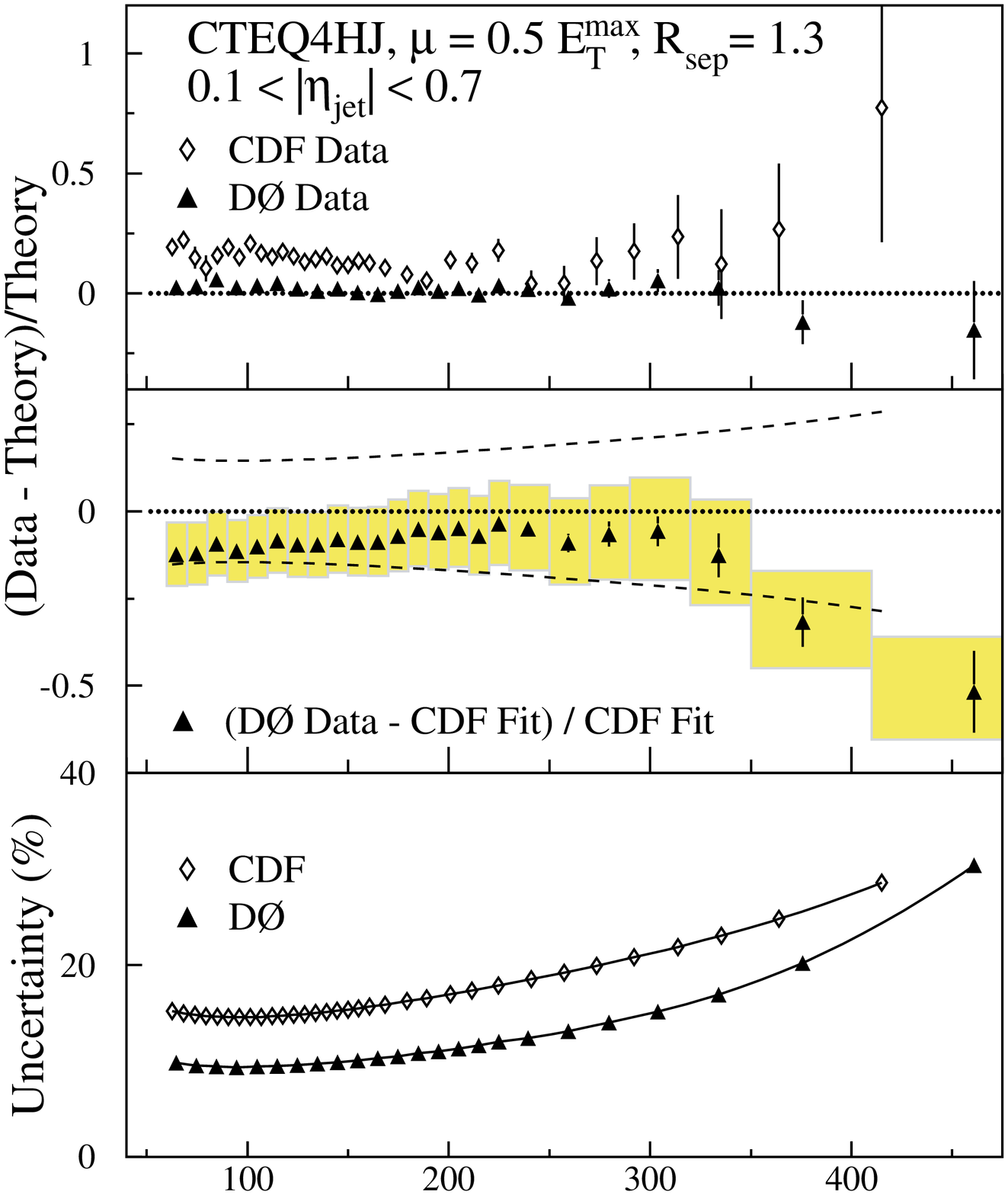}
             \end{picture}}
  \put(120,61)
    {\makebox(0,0)[rb]{\scalebox{0.6}{ \bf (c) }}}
  \put(123,65)
    {\makebox(0,0)[lt]{
      \scalebox{0.6}{ \bf \dzero\ / CDF Comparison }}}
  \put(138,55.5)
    {\makebox(0,0)[lt]{
      \scalebox{0.5}{ \bf JETRAD }}}
  \put(124,30)
    {\makebox(0,0)[lt]{
      \scalebox{0.55}{ \boldmath $\chi^{2}/\mathrm{ndf} = 32.1/24$ }}}
  \put(116,40)
    {\makebox(0,0)[lt]{
      \scalebox{0.4}{ \bf CDF Syst }}}
  \put(116,31.5)
    {\makebox(0,0)[lt]{
      \scalebox{0.4}{ \bf \dzero\ Syst }}}
  \put(128,22)
    {\makebox(0,0)[lt]{
      \scalebox{0.5}{ \bf Systematic Errors }}}
  \put(160,0)
    {\makebox(0,0)[rb]{
      \scalebox{0.6}{ \boldmath  $\et\ \mathrm{(GeV)}$}}}

  \end{picture}
  \caption{Comparisons of the \dzero\ (a) and the CDF (b) measurements of the
	   central inclusive jet cross section with theoretical predictions.
	   The error bars are statistical, while the error bands 
	   indicate systematic uncertainties (the latter not
	   presented in the CDF plot).
	   Different \peta\ regions and different  
	   calculations (JETRAD and EKS) are used in the two experiments.
	   A comparison of the two experimental results with the {\sl same} 
	   theoretical prediction (JETRAD), as well as a direct comparison 
	   between the two data sets, is shown in (c).}
  \label{fig:central}
\end{figure}

\dzero\ and CDF previously measured inclusive jet
cross sections at central \peta\ values of $\aeta<0.5$
and $0.1\leq\aeta<0.7$, respectively.
The comparisons on a linear scale between the \dzero\ 
measurement in the central $\aeta<0.5$ region and theoretical 
predictions with various PDFs are shown in 
Fig.~\ref{fig:central}a.
Furthermore, the quantitative test of agreement between
data and theory has been devised based on a $\chi^{2}$
statistic using the full covariance matrix of 
experimental uncertainties, thereby accounting for 
correlations in \et\ among different sources of error.
The $\chi^{2}$ values for the $\aeta<0.5$, and various PDFs 
used in calculations,
are presented in the Table in Fig.~\ref{fig:rapdep}c.
For purposes of comparison with the CDF measurement,
\dzero\ has also measured the inclusive jet cross
section in the $0.1\leq\aeta<0.7$ interval, and the corresponding
$\chi^{2}$ values are also summarized in the right hand
column of the same Table.
Although CTEQ4HJ PDF shows best agreement with the 
measurement, agreement with most other PDFs is also
acceptable.

CDF compares its inclusive cross section in the 
$0.1\leq\aeta<0.7$ interval to predictions from EKS 
(Ellis, \etal~\cite{theory}), with slightly modified input 
parameters.
These comparisons are presented on a linear scale in 
Fig.~\ref{fig:central}b, showing only statistical errors.
Good agreement is observed between data and theory when 
systematic experimental uncertainties are included.
Finally, at the top of Fig.~\ref{fig:central}c is shown a
comparison of CDF and \dzero\ central inclusive jet cross sections 
to the {\sl same} predictions generated using JETRAD 
with CTEQ4HJ.
The direct comparison of the two measurements is shown
in the middle plot, while the bottom plot gives the size of the 
systematic uncertainties in the CDF and \dzero\ results.
Adding the fitted CDF systematic errors in quadrature to the
\dzero\ covariance matrix, and using this matrix to calculate
the $\chi^{2}$ of agreement between the two data sets, yields 
$\chi^{2} = 32.1$ for the $24$ degrees of freedom;
This corresponds to about $12\%$ of probability---a reasonable 
level of agreement, especially given the different experimental 
techniques employed in the two measurements.

\begin{figure}[!t] \centering
  \begin{picture}(160,65)   

  \put(15.5,-1.3){\begin{picture}(70,75)
               \epsfxsize=6.7cm
               \epsfbox{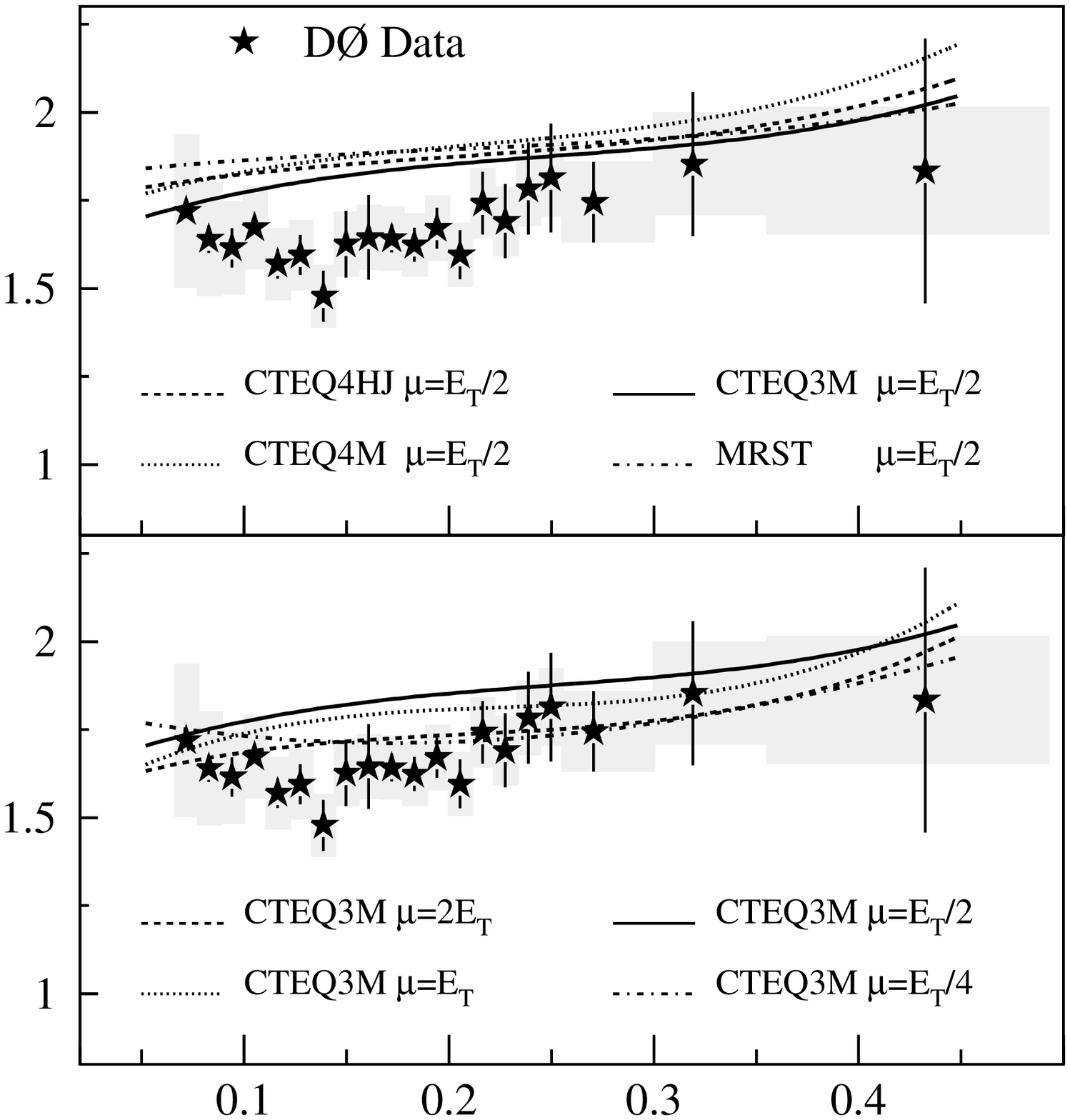}
            \end{picture}}
  \put(11.5,16){
    \scalebox{0.7}{ \rotatebox{90}{\bf Ratio of Scaled Cross Sections}}}
  \put(15,18){
    \scalebox{0.6}{ \rotatebox{90}
       {\boldmath $\left( \sqrt{s}=0.63\;\mathrm{TeV}/\sqrt{s}=1.8\;\mathrm{TeV} \right)$}}}
  \put(81.5,0)
    {\makebox(0,0)[rb]{
      \scalebox{0.6}{ \boldmath  $\mathrm{Jet\; X_{T}}$}}}
  \put(31.9,60)
    {\makebox(0,0)[rb]{\scalebox{0.6}{ \bf (a) }}}
  \put(52.5,46)
    {\makebox(0,0)[lb]{\scalebox{0.7}{ \bf Vary PDF }}}
  \put(52.5,16.5)
    {\makebox(0,0)[lb]{\scalebox{0.7}{ {\bf Vary} \boldmath $\mu${\bf-scale} }}}

  \put(80.5,3.5){\begin{picture}(70,75)
               \epsfxsize=6.01cm
               \epsfbox{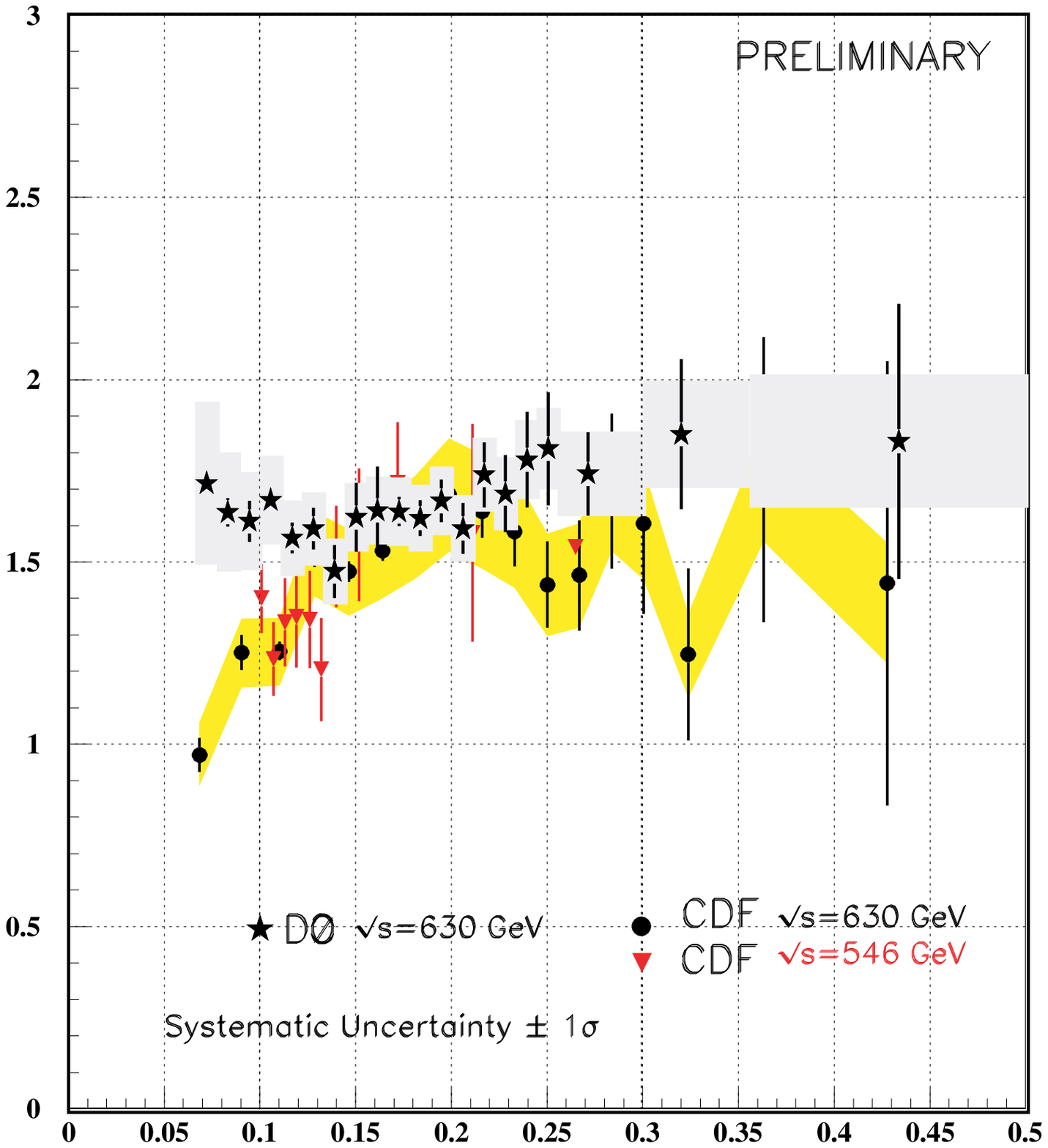}
             \end{picture}}
  \put(140.5,0)
    {\makebox(0,0)[rb]{
      \scalebox{0.6}{ \boldmath  $\mathrm{Jet\; X_{T}}$}}}
  \put(93.7,60)
    {\makebox(0,0)[rb]{\scalebox{0.6}{ \bf (b) }}}

  \end{picture}
  \caption{\dzero\ measurement of the ratio of central inclusive
	   jet cross sections from two center-of-mass energies, $0.63$ TeV
	   and $1.8$ TeV, 
	   along with predictions obtained from JETRAD (a).
	   The error bars are statistical, while the error 
	   bands indicate systematic uncertainties.	   
	   CDF measurements of the ratios corresponding to two slightly 
	   different low center-of-mass energies overlaid with the \dzero\ 
	   measurement (b).}
  \label{fig:ratio}
\end{figure}

\section{The Ratio of Inclusive Jet Cross Sections}

\noindent
\dzero\ and CDF Collaborations have recently measured the 
dimensionless ratio of inclusive jet cross sections at two 
center-of-mass energies, $\sqrt{s}=0.63$ TeV and $1.8$ 
TeV, in the central region of pseudorapidity.
The strength of this measurement is that several theoretical 
uncertainties (notably due the choice of various PDFs)
are reduced significantly in the ratio, as are many experimental
uncertainties due to their correlated nature at the two energies.
Figure~\ref{fig:ratio}a presents the \dzero\ measurement of
the ratio as a function of jet $x_{T} = 2\et/\sqrt{s}$, along
with theoretical predictions from JETRAD for different choices
of the input parameters.
Good agreement between theory and data is observed in the shape,
and the normalization appears to be in agreement within $1$--$2$
standard deviations.
The measurement of the ratio made by the CDF Collaboration
is shown in Fig.~\ref{fig:ratio}b with \dzero\ data points
overlaid to facilitate visual comparison of the two measurements.
The data sets from two experiments are qualitatively 
consistent at mid and high values of jet $x_{T}$.
At low $x_{T}$, the measurement is more difficult, and there are 
theoretical issues that could lead to disagreement with data, as 
well as between experiments.
Phenomenological choices can be made that provide better agreement
with the data.
Work is underway to obtain a quantitative measure of agreement 
between the measurements and the predictions.

%
%
%

\section*{References}

\end{document}